\documentclass[10pt]{article}
\usepackage{amsmath}
\usepackage{amsfonts,amsbsy}
\usepackage{amssymb}
\usepackage{graphicx}

\def\empile#1\over#2{\mathrel{\mathop{\kern 0pt#1}\limits_{#2}}}

\def\beq{\begin{equation}}
\def\eeq{\end{equation}}
\def\bea{\begin{eqnarray}}
\def\eea{\end{eqnarray}}

\def\p{{\boldsymbol p}}

\def\d3p{\frac{d^3\p}{(2\pi)^3}E_\p}

\catcode`\@=11

\newcount\@tempcntc
\def\@citex[#1]#2{\if@filesw\immediate\write\@auxout{\string\citation{#2}}\fi
  \@tempcnta\z@\@tempcntb\m@ne\def\@citea{}\@cite{%
        \@for\@citeb:=#2\do%
    {\@ifundefined{b@\@citeb}%
        {\@citeo\@tempcntb\m@ne\@citea%
                \def\@citea{,\penalty\@m\ }{\bf ?}\@warning%
                {Citation `\@citeb' on page \thepage \space undefined}}%
        {\setbox\z@\hbox{\global\@tempcntc0\csname b@\@citeb\endcsname\relax}
     \ifnum\@tempcntc=\z@ \@citeo\@tempcntb\m@ne%
       \@citea\def\@citea{,\penalty\@m}%
       \hbox{\csname b@\@citeb\endcsname}%
     \else%
      \advance\@tempcntb\@ne%
      \ifnum\@tempcntb=\@tempcntc%
      \else\advance\@tempcntb\m@ne\@citeo%
      \@tempcnta\@tempcntc\@tempcntb\@tempcntc\fi\fi}}\@citeo}{#1}}%

\def\@citeo{\ifnum\@tempcnta>\@tempcntb\else\@citea
  \def\@citea{,\penalty\@m}%
  \ifnum\@tempcnta=\@tempcntb\the\@tempcnta\else
   {\advance\@tempcnta\@ne\ifnum\@tempcnta=\@tempcntb \else
\def\@citea{--}\fi
    \advance\@tempcnta\m@ne\the\@tempcnta\@citea\the\@tempcntb}\fi\fi}

\catcode`\@=12

\begin{document}

\title{\bf The Electroweak Axion, Dark Energy, Inflation and Baryonic Matter\footnote{To be published in a special edition of the Journal of Experimental and Theoretical Physics in honor of the 60'th birthday of Valery Rubakov} }
\author{Larry McLerran$^{(1,2,3)}$ }

\maketitle

\begin{enumerate}

\item Physics Department, 
Brookhaven National Laboratory,
   Upton, NY 11973, USA
 \item RIKEN BNL Research Center, 
 Brookhaven National Laboratory,
   Upton, NY 11973, USA
   
   \item Physics Dept. Central China Normal University, Wuhan, China
\end{enumerate}

\begin{abstract}
In a previous paper, the standard model was generalized to include an electroweak axion which carries baryon plus lepton number, $B+L$.  It was shown that such a model
naturally gives the observed value of the dark energy, if the scale of explicit baryon number violation, $\Lambda$, was chosen to be of the order of the Planck mass.  In this paper, we consider the effect of the modulus of the axion field.  Such a field must condense in order to generate the standard Goldstone
boson associated with the phase of the axion field.  This condensation breaks baryon number.  We argue
that this modulus might be associated with inflation.  If an additional $B-L$ violating scalar is introduced with a mass similar to that of the modulus of the axion field, we argue that decays of particles associated with this field might generate an acceptable baryon asymmetry.

 \end{abstract}

\section{Introduction}

In a previous paper,  the standard model was modified by assuming that  baryon plus lepton number, $B+L$,
was not conserved at a mass scale of order the Planck scale, $\Lambda \sim M_{pl}$\cite{McLerran:2012mm}, and instantons were used
to compute a phenomenologically acceptable value of the dark energy\cite{'tHooft:1976fv}-\cite{Anselm:1993uj}.  This might be done by an electroweak axion coupling to the topological
charge of the electroweak gauge theory\cite{Peccei:1977hh}-\cite{Perez:2014fja}. Following Anselm 
and Johansen~\cite{Anselm:1993uj},  an explicit $B+L$ violating interaction of the form
\begin{equation}
 S_{B+L} = {1 \over \Lambda^2}~\int~ d^4x \left\{ \lambda l q q q  + c.c. \right\} . 
\end{equation}
was considered.
Here $l$ is a left handed lepton field and $q$ is a left handed quark field.  The scale $\Lambda$ is the energy scale at which
lepton and baryon number changing interactions are important, and is presumably a GUT scale or higher.  The matrix $\lambda$ is of order $1$, and the interaction $lqqq$ contracts various spinor, color and flavor indices to make singlets.  This interaction violates both $B+L$
and chirality.

In  a gauge theory, the $\theta$ angle appears when one considers adding a term
\begin{equation}
 n_f \theta  {\alpha \over {8\pi}}  \int~d^4x~ F \tilde{F} 
\end{equation}
to the action of the theory.  The number of families of quarks and leptons is $n_f$.  
Here $\tilde{F}_{\mu \nu} = {1 \over 2} \epsilon_{\mu \nu \lambda \sigma} F^{\lambda \sigma}$.
The quantity
\begin{equation}
{\alpha \over {8\pi}}  \int~d^4x~ F \tilde{F} = N, 
\end{equation}
where $N$ is the winding number of a Euclidean field configuration~\cite{'tHooft:1976fv,Jackiw:1976pf,Coleman:1978ae}.   Finite action solutions, instantons, exist with 
$N$ equal to the number of intstanton minus anti-instanton configurations.  The electroweak axion is generated by promoting the angle $\theta$ to an axion field.

In a theory which explicitly conserves baryon number, physics is independent of $\theta$.  In such a theory, the term above generates no dependence on $\theta$ because
the only place where instantons contribute are in amplitudes connecting states with differing numbers of baryons~\cite{'tHooft:1976fv,Krasnikov:1978dg}, and there $\theta$ 
appears as $e^{in_fN\theta}$ for a process that changes $B+L$ by amount $\Delta(B+L) = 2n_fN$.  The factor of  $n_f$ appears because each generation of quark and lepton is produced.  The basic instanton process therefore involves 9 colored quarks and three leptons.  In amplitudes squared, the phase disappears and there is no consequence of this angle.

It was shown in the previous paper that if there was explicit baryon number violation at the Planck scale, then electroweak instanton processes naturally led to a vacuum energy of the magnitude
\begin{equation}
           S_{\rm I}  = \kappa (cos(n_f \theta) -1) \left( {{2\pi} \over {\alpha_{\rm W}}} \right)^4 \left( {M_{\rm EW} \over \Lambda} \right)^{19/6}  e^{-2\pi/\alpha_{\rm W}(M_{\rm EW})}  \Lambda^4. 
\end{equation}

If we take the energy scale $\Lambda$ to be the Planck mass, and $1/\alpha_{\rm W} \sim 1/30$, we find that
\begin{equation}
           S_{\rm I} \sim 10^{-122} \cdot M_{\rm pl}^4. 
\end{equation}
This is remarkably close to the value of dark energy in cosmology, $\epsilon_{\rm DE} \sim 10^{-122} \cdot M_{\rm pl}^4 $.

There are of course uncertainties in this estimation of the scale $\Lambda$.  The details of $B+L$ violation may be different
in different theories, and there may be some changes to the coupling contsant evolution at energies near the Planck
scale due to new particle degrees of freedom.  The formula for the dark energy is roughly linear in $\Lambda$, so that 
an uncertainty in this relationship translates to a roughly linear uncertainty in the scale $\Lambda$. This linear dependence arises from the explicit factors of $\Lambda^4$ and the implicit factors in the running of the coupling constant. It is not unreasonable to assume that there may be several orders of magnitude uncertainty in the scale $\Lambda$, since the running of the coupling constant is not known near the Planck scale.

\section{Generalization to the Non-Goldstone Mode}

The axion field above is the Goldstone mode composed from fields $\phi = \rho(x) e^{i\theta(x)}$. 
We will assume that the field $\phi$ carries one unit of $B+L$.  (One unit of B+L corresponds to $B+L = 2$). The anomaly however generates $n_f$ units
of $B+L$ change, coprrespodnding to $\Delta(B+L) = 2n_f$.  
 The Lagrangean with only axion degrees of freedom arises when we assume that a symmetry associated with scalar field is spontaneously broken and the field $\rho$ acquires an
 expectation value.  We should think of the term $cos(n_f \theta)$ in the induced instanton interaction of Eqn. 4 as
 \begin{equation}
  cos(\theta) = {1 \over 2}(e^{in_f\theta} + e^{-in_f\theta}) = Re(\phi^3)/v^3
 \end{equation}
 where 
 \begin{equation}
  v = <\rho>
 \end{equation}
 When the field is replaced by its expectation value then we achieve our old result.  
 
 The contribution of the axion to the action is very small.  However, multiple weak boson attatchments to the basic vertex can enhance the magnitude of B+L violation, and at high temperatures, $1/\alpha_W$ such enhancements make the effect of  magnitude sufficent for the processes to be realizable\cite{Kuzmin:1985mm}-\cite{Arnold:1987zg}.  One might ask if the contribution associated with the dynamical non-Goldstone part of the axion field might be similarly enhanced, for example in the decay of a heavy axion.  We think not since the axion brings an energy scale into the problem much larger than the electroweak scale, and the amplitude for such a decay enhanced by thermal W and Z bsoson should maintain its exponential suppresion $\sim e^{-2\pi/\alpha_W}$.

In Ref. \cite{McLerran:2012mm},
it was assumed that the symmetry was broken, that $ v$ had  an expectation value of order $M_{pl}$,
and that $\theta(x)$ was frozen into a constant value by inflation.  In fact, the modulus of the axion field,
$\rho$ provides a candidate for the inflaton field\cite{Wetterich:1987fm}.  It has an expectation value of order the Planck mass,
as is required of the inflaton in some inflationary scenarios \cite{Starobinsky:1980te}-\cite{Albrecht:1982wi}.  
In order to get the right order of magnitude for density fluctuations,\cite{Starobinsky:1980te}-\cite{Bardeen:1983qw}  we will need to require a very flat potential for the modulus of the axion field.  This would require mass $m_{\rho} << m_{pl}$  for the scalar particle associated with the modulus of the axion mass.    A typical value for the inflaton mass in chaotic inflation scenarios is $10^{12}~GeV$.

As the symmetry breaking occurs, the $B+L$ symmetry is spontaneously broken, and it is plausible that some excess of the heavy scalar particles associated with the non-Goldstone part of the scalar field are produced.  These scalar particles
will carry non-zero baryon number.  They will have B+L violating interactions among themselves but when the density of such particles becomes sufficiently low, we expect these interactions will freeze out.  However,
these particles decay rapidly because in the axion action after symmetry breaking, there is a term
\begin{equation}
  \delta S = \int~d^4x~ v~\delta \rho~ \partial^\mu \theta \partial_\mu \theta
\end{equation}
that allows the modulus of the scalar field to decay into two axion fields.

If we further assume there is a $B-L$ violating scalar with scale of variation at the Planck mass and a mass similar to that of the modulus of the $B+L$ violating modulus of the scalar field, then there may be interesting effects.  Lets us assume that there is a $B-L$ symmetry of the scalar field action, but there are $B-L$ violating interactions with quarks and leptons.  We will see that these interactions with quarks and leptons are quite weak at low energy scales.  Then it is plausible that at a high energy scale associated with the end of inflation, one might generate some excess of $B-L$ which is stored in the low mass $m_{B-L} << M_{pl}$ scalar field.   If this is the case, then such matter plays an increasingly important role as time evolves, since massive matter energy density dilutes more slowly than does radiation.
At some late time, it is not implausible to assume that such matter dominates the energy density of the universe.  

There may be baryon number violating processes where the massive $B-L$ scalar field would decay into light mass quarks and leptons.  On dimensional grounds, the effective interaction for such a term is 
\begin{equation}
              L_{eff} = {1 \over \Lambda^3} \sum_i \phi_{B-L} \overline{l}_i q_iq_iq_i
\end{equation}
where the sum is over quark $q$ and lepton $l$ flavors.  The parameter $\Lambda$ is of order the Planck mass.  There is still a considerable uncertainty in the value of $\Lambda$.  In the derivation of the vacuum energy, the running of the electroweak coupling and its dependence upon the Planck mass scale combine with the explict factors of $M_{pl}^4$ to make for an almost linear sentitvity of the dependence of the instanton induced vacuum energy on the Planck mass.  This, combined with the intrinsic uncertainty of how the electroweak coupling runs at energy scales near the Planck mass allows for a few orders of magnitude
uncertainty upon the energy scale at which baryon number violation is of order 1.

The rate for decay is
\begin{equation}
    R \sim m_{B-L}^7/\Lambda^6 
    \label{rate1}
\end{equation}
which is anomalously small because $\Lambda/m_{B-L} \sim 10^7$.  For example, if expansion was radiation dominated,  which is what we want to match to as the matter reheats, then this rate would become equal to the expansion rate when
\begin{equation}
   T \sim M_{pl} (m_{B-L}/\Lambda)^{7/2}
\end{equation}

Now in order for the decay of the scalar field not to produce too many baryons,  it is necessary that the decaying baryon not produce too little entropy.  If the reheating temperature is $T$, there will be of order
$m_{B-L}/T$ particles produced per unit baryon number.  For a scalar mass of order $m_{B-L} \sim 10^{12}~ GeV$ this would require $T \ge 100~GeV$.  If the temperature is signifcantly above $100~GeV$, then any asymmetry generated by the decays of such bosons is preserved by sphaleron decays,
since these decay only violate $B+L$\cite{Kuzmin:1985mm}.  If the temperature is near to  $100~GeV$ then we would generate an acceptable baryon asymmetry.  So we see that for the mass scale of order $10^{12}~GeV$ there is some  narrow temperature range where
one can make an acceptable baryon asymmetry.   Outside of this temperature range either there is either too much or too little baryon asymmetry. 

If we use Eqn. \ref{rate1} to  get a reheat temperature around $100~GeV$ we would require $m_{B-L} \sim 10^{-5} \Lambda$. If $\Lambda$ was the Planck mass, this would require a  mass of $10^{14}~GeV$, which is large compared to the  expected value of the inflaton mass.    This mass scale would generate an acceptable asymmetry if the reheat temperature is $10~TeV$.

We should however recall that the remaining particles left after the inflationary transition is accomplished
are scalar particles.  In numerical simulations of the evolution of an over-occupied scalar field, one always forms a condensate \cite{Micha:2002ey}-\cite{Berges:2013eia}.  Over-occupation might  generally be a good starting assumption if the scalar particles arise from a coherent scalar field.  
In general, for a scalar field we would expect a transient condensate to form associated with the scalar fields as the system expands.  This condensate would oscillate in time, but have zero spatial momentum.  

Scalar bosons always have an attractive energy associated interactions.  Also the range of interactions is very large, of order $1/m $, which may be quite long compared to event horizon size scales when the
the inflaton field begins condensation. Therefore in this condensation, it is not implausible that as the universe expands, the condensate breaks up into 
large regions of clustered scalar particles that have a coherent field of the order 
$\phi_{B-L} \sim m_{B-L}/\sqrt{\lambda}$.  It would be most intersting to find an explict scenario where such $q-balls$ exist, and perform simulations to determine whether such a scenario is indeed plausible.
We will  assume q-balls somehow form\cite{Coleman:1985ki}-\cite{Kusenko:1997ad}.  Eventually,
even if the energy density of such q-balls was small campared to the energy density stored in radiation,
the q-ball energy density would eventually dominate the energy density of the universe.

If the scalar bosons are condensed, we expect that their occupation number would be of the order of $1/\lambda$ where $\lambda$ is the magnitude of some effective scalar four point interaction.  This should be of the order
\begin{equation}
\lambda \sim m_{B-L}^2/\Lambda^2
\end{equation}
The decay rate formual would be enhanced by a factor of $1/\lambda$ so our paramtertrc estimate of the decay rate is relapced by
\begin{equation}
R \sim m_{B-L}^5/\Lambda^4 
\end{equation}
to that the time when the condensate dcays is
\begin{equation}
T \sim M_{pl} (m_{B-L}/\Lambda)^{5/2}
 \end{equation} 
 If we take $\Lambda \sim M_{pl}$, and $m_{B-L} \sim 10^{12}~GeV$, we naturally get a reheating temperature of order $T \sim 100~GeV$.

\section{Summary}

We have argued that an electroweak axion may in principle have the correct dynamics to generate inflation,
Including an extra $B-L$ violating scalar with mass similar to the mass of the modulus of the axion field gives
an accptable baryon asymmetry, The scales one introduces in order to make this consistent with what is known from cosmology are natural.    

The computation we have done follows the philosophy of Shaposhnikov and Wetterich\cite{Shaposhnikov:2009pv}, 
that one should push the limit of  of the standard model as far as possible making only minimal changes to its structure to include new physics.  
The picture we paint is somehwat similar to that of Affleck and Dine\cite{Affleck:1984fy}. as far as the baryon number generation is concerned,
and indeed it would be interesting to find a supersymmetric derivation of an action which has the properties we need
to get an acceptabl dark energy and baryon number density.  Perhaps such a generalization would lead naturally
to an explanation of dark matter as well.  The framework of Shaposhnikov and  colleagues for the neutrino sector and 
standard model cosmology may be applicable here\cite{Bezrukov:2007ep}. 


\section*{Acknowledgements}
 We thank Rob Pisarski and Hooman Davioudasl for enlightening discussions. 
 The research L. McLerran  is supported under DOE Contract No. DE-AC02-98CH10886. 
 This work was completed while L. McLerran was visitng the Theoretical Physics Institute at the University of Heidelberg as the Jensen Professor of Theoretical Physics.

\end{document}